\documentclass[pre,twocolumn,showpacs]{revtex4}     
\usepackage{epsfig,amsmath} 

\begin{document}

\title{Approximation for discrete Fourier transform and application in study of three-dimensional interacting electron gas} 

\author{Xin-Zhong Yan}
\affiliation{Institute of Physics, Chinese Academy of Sciences, P.O. Box 603, Beijing 100190, China}
 
\date{\today}
 
\begin{abstract}
The discrete Fourier transform is approximated by summing over part of the terms with corresponding weights. The approximation reduces significantly the requirement for computer memory storage and enhances the numerical computation efficiency with several orders without loosing accuracy. As an example, we apply the algorithm to study the three-dimensional interacting electron gas under the renormalized-ring-diagram approximation where the Green's function needs to be self-consistently solved. We present the results for the chemical potential, compressibility, free energy, entropy, and specific heat of the system. The ground-state energy obtained by the present calculation is compared with the existing results of Monte Carlo simulation and random-phase approximation.   
\end{abstract}


\maketitle

\section{Introduction}

For dealing with some physical problems, we need to take the discrete Fourier transform. Especially, many physical problems are defined on lattice models. In such a case, we may face to the problem of Fourier transforming a function defined on the lattice to the corresponding reciprocal space. For most of the problems, the function to be transformed cannot be expressed analytically but given numerically. When the function needs to be given at a large number of discrete points within the region it is defined, the memory volume for storing the function may be too big and may even exceed the computer's storage limit. Even if the problem is within the computer's capability, when the transform is involved in an integral equation that may be solved by iterations, the function needs to be determined again and again in the iterations and the process is very time consuming. Therefore, an approximation scheme for the discrete Fourier transform that reduces the storage requirement and accelerates the numerical computation process without loosing the accuracy is very desirable.

The discrete Fourier transform as well as a continuous one is useful in solving the integral equations with convolutions involved. One of the examples in the quantum many-body problems is to calculate the self-energy $\Sigma$ of electrons \cite{Baym,Baym2},
\begin{eqnarray}
\Sigma(k,i\omega_n) = -\frac{T}{V}\sum_{k'n'}v_{eff}(k-k',i\omega_n-i\omega_{n'})G(k',i\omega_{n'}) \nonumber\\
\label{ise}
\end{eqnarray}
where $k$ is the momentum of electron, $\omega_n = \pi T(2n+1)$ with $n$ an integer is the fermionic Matsubara frequency, $T$ is temperature, $V$ is the volume of the system, $v_{eff}$ is an effective interaction between electrons, and $G$ is the Green's function of electrons. In a sophisticated scheme, $\Sigma$ and $G$ need to be determined self-consistently. After the Fourier transforms, in real coordinate $r$ and imaginary time $\tau$ space, Eq. (\ref{ise}) reads 
\begin{eqnarray}
\Sigma(r,\tau) = -v_{eff}(r,\tau)G(r,\tau). \label{ise2}
\end{eqnarray}
Having $\Sigma(r,\tau)$ been simply calculated by Eq. (\ref{ise2}), one then obtains $\Sigma(k,i\omega_n)$ by the inverse Fourier transforms. 

Earlier works dealt with Eq. (\ref{ise}) by direct summation over the Matsumara frequencies. In order to reduce the memory storage requirement and accelerate the computation process, Pao and Bickers developed a renormalization-group computation method \cite{Pao}. The method is based on the assumption that the Green's function depends approximately on $T$ only through the Matsubara frequency. The computation starts at high temperature $T_0$ to solve the equations of the Green's function at selected numbers $\{n\}$ with cutoff $N_0$ for the Matsubara frequencies. Since the Green's function decreases with Matsubara frequency very fast at high temperature, the selected numbers is not necessarily too many. Then at lower temperature $T_1$, the selected numbers correspond to lower frequencies. The equations for the functions at these lower frequencies are solved and the functions at some high frequencies $\omega_n > \omega_{N_0}$ are approximated as the ones calculated at $T_0$. For example, $G(k,i\omega_n)_{T_1,n>N_0} \approx G(k,i\omega_n')_{T_0}$ with $\omega_n = \pi T_1(2n+1) = \pi T_0 (2n'+1)$ where $n'$ is the selected number. The series summation in Eq. (\ref{ise}) is carried out using the staircase rule. That is, summation of $f(n)$ over the range $n_1 \le n \le n_2-1$ with $n_1$ and $n_2$ the any two nearest-neighbor selected numbers is given by $f(n_1)(n_2-n_1)$. By repeating this sequence, the equations for determining the Green's function are so solved at lower temperatures.  

The key problem in solving such integral equations by the direct summation treatment is how to accurately take the series summation with the selected numbers. For numerically computing the series summation, 
\begin{eqnarray}
S = \sum_{n=0}^{\infty}f(n), \label{ism}
\end{eqnarray}
the present author has introduced an algorithm that sums over selected numbers with corresponding weights. The basic idea of this method is described as following. Suppose $f$ as a function of continuously variable $x$ is locally smooth. Between the selected numbers $n_1$ and $n_3$ with $n_3 = n_1 + 2h$ and $h$ an integer, $f(n)$ can be expanded as
\begin{eqnarray}
f(n) \approx f(n_1) + c_1(n-n_1) + c_2(n-n_1)^2 \label{fj}
\end{eqnarray}
where the coefficients $c_1$ and $c_2$ are determined by the function values $f(n_2)$ with $n_2 = n_1 + h$ (the midpoint between $n_1$ and $n_3$ also selected) and $f(n_3)$. They are given by
\begin{eqnarray}
c_1 &=& [-3f(n_1)+4f(n_2) -f(n_3)]/2h, \label{c1}\\
c_2 &=& [  f(n_1)-2f(n_2) +f(n_3)]/2h^2. \label{c2}
\end{eqnarray}
Then using the results, 
\begin{eqnarray}
\sum_{j=1}^n k &=& n(n+1)/2, \label{sm1}\\
\sum_{j=1}^n k^2 &=& n(n+1)(2n+1)/6, \label{sm2}
\end{eqnarray}
the summation of $f(n)$ over the range $n_1 \le n \le n_3-1$ is obtained approximately in terms of $f(n_1)$, $f(n_2)$, and $f(n_3)$ the values of $f$ all at the selected points; the coefficients attached respectively to these values are the corresponding weights depending only on $h$. By repeating this procedure to a large cutoff number, the summation in Eq. (\ref{ism}) is then obtained. The algorithm is proved to be very accurate.

Though the convolution with the discrete numbers can be treated as series summation, numerical computation with the discrete Fourier transform is much easier. It is even faster provided the transform is performed using a high efficiency algorithm. In this work, we will develop an algorithm to the discrete Fourier transform. The accuracy and efficiency of the new algorithm will be justified with examples. 

In the later part of this paper, we will apply the algorithm to the physical problem studying three-dimensional interacting electron gas (3DEG) under the renormalized-ring-diagram approximation (RRDA) \cite{Yan2} and compare the ground-state energy so obtained with existing results of the Monte Carlo (MC) simulation \cite{Ceperley} and the random-phase approximation (RPA). RRDA satisfies the microscopic conservation laws \cite{Baym,Baym2}. It has not so far been applied to 3DEG because of the numerical difficulty in self-consistently solving the integral equations determining the Green's function.  

\section{approximation for the discrete Fourier transform} 

We here consider the discrete Fourier transform
\begin{eqnarray}
F(k) = \sum_{j=n_a}^{n_b}f(j)\exp(-ikj) \label{sum}
\end{eqnarray}
where $f(j)$ is defined in the range $n_a \le j \le n_b$ with $n_a$ and $n_b$ being integer numbers and $k$ is a real parameter in the range $(-\pi,\pi)$. To find out an approximation for it, we firstly analyze the following summation in small range $(n_1,n_3)$ with $n_3 - n_1 = 2h$ and $n_1$, $n_3$ and $h$ all integers, 
\begin{eqnarray}
F(n_1,n_3;k) = \sum_{j=n_1}^{n_3-1}f(j)\exp(-ikj). \label{sum}
\end{eqnarray}
For large $k$, since $\exp(-ikj)$ is a rapid oscillating factor, $f(j)\exp(-ikj)$ cannot be regarded as a smooth function of $j$ and the previous algorithm cannot be applied here. However, for smooth function $f(x)$ in the range $n_1 < x < n_3$, $f(j)$ can be expanded as in Eq. (\ref{fj}). We can then obtain an approximated result for $F(n_1,n_3;k)$. We need the following summation 
\begin{eqnarray}
S_1(k) &=& \sum_{j=n_1}^{n_3-1}\exp(-ikj) \nonumber\\
&=&  \exp(-ikn_1)\frac{1-\exp(-i2kh)}{1-\exp(-ik)}  \nonumber\\
&\equiv &  \exp(-ikn_1)y(k)            \label{s1}
\end{eqnarray}
with $y(k) = [1-\exp(-i2kh)]/[1-\exp(-ik)]$. Then we have  
\begin{eqnarray}
S_2(k) &=& \sum_{j=n_1}^{n_3-1}(j-n_1)\exp(-ikj) \nonumber\\
&=& i\exp(-ikn_1)dy(k)/dk \label{s2}\\
S_3(k) &=& \sum_{j=n_1}^{n_3-1}(j-n_1)^2\exp(-ikj) \nonumber\\
&=& -\exp(-ikn_1)d^2y(k)/dk^2. \label{s3}
\end{eqnarray}
Substituting Eqs. (\ref{fj}) and (\ref{s1})-(\ref{s3}) into Eq. (\ref{sum}), we get
\begin{eqnarray}
F(n_1,n_3;k) \approx f(n_1)S_1(k)+c_1S_2(k)+c_2S_3(k).     \label{sm1}
\end{eqnarray}
Using Eqs. (\ref{c1}) and (\ref{c2}), we obtain
\begin{eqnarray}
F(n_1,n_3;k) &\approx & w_1(k)f(n_1)\exp(-ikn_1)\nonumber\\
  &&+w_2(k)f(n_2)\exp(-ikn_2)\nonumber\\
  &&+w_3(k)f(n_3)\exp(-ikn_3) \label{sm2}
\end{eqnarray}
where the weight functions $w_{1,2,3}(k)$ are given by
\begin{eqnarray}
w_1(k) &=& y(k) -i\frac{3}{2h}\frac{dy(k)}{dk} - \frac{1}{2h^2}\frac{d^2y(k)}{dk^2}, \nonumber\\
w_2(k) &=& \exp(ikh)[i\frac{2}{h}\frac{dy(k)}{dk} + \frac{1}{h^2}\frac{d^2y(k)}{dk^2}], \nonumber\\
w_3(k) &=& \exp(i2kh)[-\frac{i}{2h}\frac{dy(k)}{dk} - \frac{1}{2h^2}\frac{d^2y(k)}{dk^2}]. \nonumber
\end{eqnarray}
Note that these weights depend only on the parameters $k$ and $h$. Equation (\ref{sm2}) means that the summation over the range $n_1 \le j \le n_3-1$ can be approximately obtained by the three values of the function at $n_1$, $n_2$ and $n_3$. 

Now, we select equal spaced numbers $(n_1,n_2,\cdots,n_{2m+1})$ with integer stride $h$ and consider the summation
\begin{eqnarray}
F(n_1,n_{2m+1};k) = \sum_{j=n_1}^{n_{2m+1}-1}f(j)\exp(-ikj). \label{tsum}
\end{eqnarray}
Using Eq. (\ref{sm2}), it can be expressed as
\begin{eqnarray}
&&F(n_1,n_{2m+1};k) = \sum_{\ell=1}^mF(n_{2\ell-1},n_{2\ell+1};k) \nonumber\\
 \approx && w_2(k)\sum_{\ell=1}^mf(n_{2\ell})e^{-ikn_{2\ell}}\nonumber\\
 &&+[w_1(k)+w_3(k)]\sum_{\ell=1}^mf(n_{2\ell-1})e^{-ikn_{2\ell-1}}\nonumber\\
 &&+w_3(k)[f(n_{2m+1})e^{-ikn_{2m+1}}-f(n_1)e^{-ikn_1}]\nonumber\\
 \equiv &&w_e(k)F_e(k)+w_o(k)F_o(k)+w_x(k)F_x(k) \label{tsum1}
\end{eqnarray}
with
\begin{eqnarray}
F_x(k)&=&f(n_{2m+1})e^{-ikn_{2m+1}}-f(n_1)e^{-ikn_1}  \label{fx}\\     
F_o(k)&=&\sum_{\ell=1}^mf(n_{2\ell-1})e^{-ikn_{2\ell-1}}+\frac{1}{2}F_x(k)\label{fo}\\ 
F_e(k)&=&\sum_{\ell=1}^mf(n_{2\ell})e^{-ikn_{2\ell}}  \label{fe}
\end{eqnarray}
and $w_e(k) = w_2(k)$, $w_o(k) = w_1(k) + w_3(k)$ and $w_x(k) = [w_3(k)-w_1(k)]/2 \equiv iw_i(k)-1/2$. Clearly, $F_o(k)$ is the summation over the odd terms with the trapezoid rule which counts half the values at the two ends, $F_e(k)$ is that over the even terms, and $F_x(k)$ is the extra term. Inserting the function $y(k)$ in the formula $w_{1,2,3}(k)$, the final expressions for $w_e(k)$, $w_o(k)$, and $w_i(k)$ are obtained as
\begin{eqnarray}
w_e(k)&=& [\frac{\sin(kh)}{h\tan(k/2)}-2\cos(kh)]/h(1-\cos k), \label{we}\\
w_o(k)&=& \frac{\sin^2(kh)}{h(1-\cos k)}-w_e(k)\cos(kh), \label{wo}\\
w_i(k)&=& \frac{h\sin k-\sin(kh)\cos(kh)}{2h(1-\cos k)}-\frac{1}{2}w_e(k)\sin(kh). \nonumber\\
\label{wi}
\end{eqnarray}
At $k = 0$, these functions take their limit values
\begin{eqnarray}
w_e(0)&=& \frac{h}{3}(4-\frac{1}{h^2}), \nonumber\\
w_o(0)&=& \frac{h}{3}(2+\frac{1}{h^2}), \nonumber\\
w_i(0)&=& 0, \nonumber
\end{eqnarray}
which are the same weights obtained previously for the series summation \cite{Yan}. The formulas (\ref{tsum1})-(\ref{wi}) are the main result here. Equation (\ref{tsum1}) is valid when the function $f(x)$ is smooth in each segment ($n_{2\ell-1}$, $n_{2\ell+1}$). As shown above, only in each of these segments, $f(x)$ is approximated as a parabolic function and then the summation over the terms with the oscillating exponential factor is performed exactly. By separating the whole range $(n_a, n_b)$ into several pieces with different strides according to the behavior of the function and using the summation rule given above in each piece, we then obtain the discrete Fourier transform. This reduces greatly the memory storage and enhances the computation efficiency. Here is a remark: Except for the number of the selected points is odd in each piece, there is no constraint on the numbers of values of the input function $f$ and the output results and no constraint on the relation between the stride of $k$ and the total number of the selected points $\{n_i\}$, which is different from the condition of the fast Fourier transform. Therefore, it is convenient for using.

Some problems may be related to the sine or cosine transform:
\begin{eqnarray}
S(k) = \sum_{j=n_1}^{n_{2m+1}-1}f(j)\sin(kj), \label{sin}\\
C(k) = \sum_{j=n_1}^{n_{2m+1}-1}f(j)\cos(kj). \label{cos}
\end{eqnarray}
By recognizing the functions $\cos(kj)$ and $\sin(kj)$ are respectively the real and negative imaginary parts of $\exp(-ikj)$, from Eq. (\ref{tsum1}), we obtain
\begin{eqnarray}
S(k) &=& w_e(k)S_e(k)+w_o(k)S_o(k)-\frac{1}{2}S_x(k)-w_i(k)C_x(k)\nonumber\\
C(k) &=& w_e(k)C_e(k)+w_o(k)C_o(k)-\frac{1}{2}C_x(k)+w_i(k)S_x(k)\nonumber
\end{eqnarray}
with
\begin{eqnarray}
S_x(k)&=&f(n_{2m+1})\sin(kn_{2m+1})-f(n_1)\sin(kn_1) \nonumber\\  
S_o(k)&=&\sum_{\ell=1}^mf(n_{2\ell-1})\sin(kn_{2\ell-1})+\frac{1}{2}S_x(k)\nonumber\\
S_e(k)&=&\sum_{\ell=1}^mf(n_{2\ell})\sin(kn_{2\ell})  \nonumber\\
C_x(k)&=&f(n_{2m+1})\cos(kn_{2m+1})-f(n_1)\cos(kn_1) \nonumber\\    
C_o(k)&=&\sum_{\ell=1}^mf(n_{2\ell-1})\cos(kn_{2\ell-1})+\frac{1}{2}C_x(k)\nonumber\\
C_e(k)&=&\sum_{\ell=1}^mf(n_{2\ell})\cos(kn_{2\ell}).  \nonumber
\end{eqnarray}

\begin{figure} 
\centerline{\epsfig{file=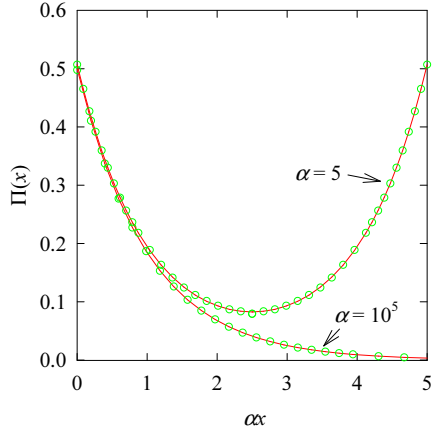,height=6.cm}}
\caption{(color online) $\Pi(x)$ as function of $x$ at parameters $p = 5$ and $p = 10^5$. Solid lines represent the exact function. Circles are the numerical results obtained using the present summation rule.} 
\end{figure} 

To test the accuracy and efficiency of the summation rule, we here consider an example,
\begin{eqnarray}
\Pi(x) = \sum_{n=-\infty}^{\infty}\frac{p}{p^2+(2\pi n)^2}\exp(-i2\pi nx)\label{ex1}
\end{eqnarray}
where $p$ is a parameter. The exact result of the summation is 
\begin{eqnarray}
\Pi(x) = \frac{1}{2}\frac{\exp(-p x)+\exp[p(x-1)]}{1-\exp(-p)}, \nonumber
\end{eqnarray}
for $0 \le x \le 1$. $\Pi(x)$ is a periodic function of $x$ with periodicity 1. For numerical calculation, because the term under summation is even with $n \to -n$, we rewrite Eq. (\ref{ex1}) as
\begin{eqnarray}
\Pi(x) = -\frac{1}{p}+2\sum_{n=0}^{\infty}\frac{p}{p^2+(2\pi n)^2}\cos(2\pi nx). \label{ex11}
\end{eqnarray}
To get the summation converged in numerical calculation, the cutoff $N_0$ that is the terms we need to sum should be much larger than $p/2\pi$. For example, supposing $p = 10^5$, we may take $N_0 = 100 p/2\pi \approx 1.6\times 10^6$. For $x \sim 0$, the contribution from the remaining term of $n > N_0$ is $O(p/2\pi^2N_0) \sim O(1/100\pi)$. According to the summation rule, instead of summing term by term within the cutoff, summation in Eq. (\ref{ex11}) is taken over only selected numbers. We here use our previous number-selection scheme \cite{Yan}: The selected numbers distribute in $L$ successively connected blocks (pieces) in the positive integer-number axis. Each block contains $M$ equal spaced numbers (selected). (Including the two ends, there are $M+1$ numbers in each block. The number $M$ is redefined here. In different from the previous notation where $M$ was defined as the total number including the two ends, here it counts the number on one end.) The stride (or the length between two selected numbers) in the $\ell$th block is $h_{\ell} = h^{\ell-1}$ with $h$ a constant integer number. For this example, we here use $[h,L,M] = [2,19,4]$. The total number of the selected numbers is $LM+1 = 77$, but the cutoff is $N_0 = Mh^L/(h-1)-(M-h+1)/(h-1) \approx Mh^L/(h-1) = 2^{21} \approx 2.1\times 10^6$. Therefore, the cutoff should be large enough for $p \le 10^5$. Figure 1 shows the present numerical results (circles) for $\Pi(x)$ as function of $x$ at two parameters $p = 5$ and $p = 10^5$. The calculation is compared with the exact formula given by the solid lines in Fig. 1. For $p =5$, Fig.1 just shows the result within a periodic range. For $p = 10^5$, $\Pi (x)$ takes sizable value when $x$ is close to 0 or 1 within the periodic range $0 < x <1$ and is symmetric about $x = 0.5$. Here we plot only the result for $x$ close to 0 for $p = 10^5$ because the scale $p$ is too large to depict the result in a complete periodic range. It is seen that the numerical calculation very accurately reproduces the exact results. As for the efficiency, because the summation of $N_0 \approx Mh^L/(h-1)$ terms is approximately obtained by summing over only $LM+1$ terms, the efficiency $c$ can be defined as $N_0/[LM+1]$. That is,
\begin{eqnarray}
c \approx h^L/L(h-1). \label{eff}
\end{eqnarray}
For the present example, the efficiency is $c = 2.7\times 10^4$. 

In dealing with a physical problem, we may face to summation like
\begin{eqnarray}
\Pi(x,p_1,p_2,\cdots) = \sum_{n=-\infty}^{\infty}f(n,p_1,p_2,\cdots)\exp(-i2\pi nx) \nonumber
\end{eqnarray}
where $f(n,p_1,p_2,\cdots)$ cannot be explicitly expressed but given numerically. $\Pi(x,p_1,p_2,\cdots)$ as function of its arguments needs to be computed numerically in the region of $(x,p_1,p_2,\cdots)$ where the summation converges. According to the present summation rule, first, we only need to numerically calculate $f$ at the selected numbers, which saves computation time and memory storage for $f$. Second, the high efficiency summation algorithm saves time significantly for getting $\Pi(x,p_1,p_2,\cdots)$. 

\begin{figure} 
\centerline{\epsfig{file=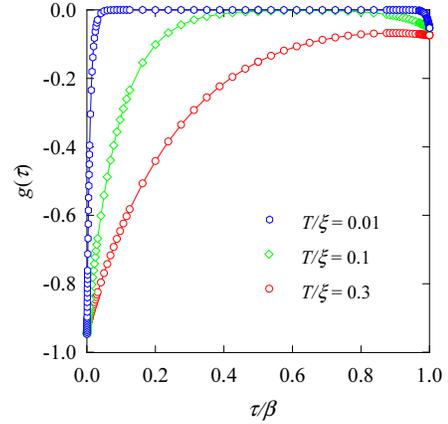,height=6.cm}}
\caption{(color online) $g(\tau)$ as function of $\tau$ (normalized by $\beta = 1/T$) at parameters $T/\xi = 0.01$ 0.1 and 0.3. Solid lines represent the exact function. Symbols are the numerical results obtained using the present summation rule.} 
\end{figure} 

To effectively apply the present summation algorithm, we here consider another example, 
\begin{eqnarray}
g(\tau) = \frac{1}{\beta}\sum_{n=-\infty}^{\infty}\frac{\exp(-i\omega_n\tau)}{i\omega_n -\xi-\Sigma(i\omega_n)} \label{ex2}
\end{eqnarray}
where $\beta = 1/T$, $\omega_n$ is the fermionic Matsubara frequency, and $\Sigma(i\omega_n) = \Delta^2/(i\omega_n+\xi)$ is the self-energy with $\Delta$ a parameter. The exact result is 
\begin{eqnarray}
g(\tau) &=& -\frac{1}{2}(1+\frac{\xi}{E})F(-E)\exp(-E\tau) \nonumber\\
&& -\frac{1}{2}(1-\frac{\xi}{E})F(E)\exp(E\tau) \nonumber
\end{eqnarray}
for $0 < \tau < \beta$, where $E = \sqrt{\xi^2+\Delta^2}$ and $F(E) = 1/[\exp(\beta E)+1]$ is the Fermi distribution function. Note that $G(i\omega_n) \equiv 1/[i\omega_n -\xi-\Sigma(i\omega_n)] \sim 1/i\omega_n$ as $n \to \infty$. Therefore, the summation in Eq. (\ref{ex2}) for $\tau \to 0$ is not absolutely converging but converges conditionally. In such a case, one usually makes use of auxiliary function to accelerate the convergence in numerical calculation. We here choose the auxiliary function as $G^0(i\omega_n) = 1/(i\omega_n -\xi)$. The summation 
\begin{eqnarray}
g^0(\tau) &=& \frac{1}{\beta}\sum_{n=-\infty}^{\infty}G^0(i\omega_n)\exp(-i\omega_n\tau) \nonumber\\
&=& -F(-\xi)\exp(-\xi\tau), ~~~~{\rm for}~0 < \tau < \beta \nonumber
\end{eqnarray}
is known. We then need to do the numerical calculation given as, 
\begin{eqnarray}
g(\tau) = \frac{2}{\beta}\sum_{n=0}^{\infty}{\rm Re}\{[G(i\omega_n)-G^0(i\omega_n)]e^{-i\omega_n\tau}\}+g^0(\tau) \nonumber
\end{eqnarray}
where the summation is absolutely converging in the limit $\tau \to 0$. Using our number-selection scheme with $[h,L,M] = [2,17,4]$, we numerically calculate $g(\tau)$. The obtained results (symbols) are shown in Fig. 2 for parameters $T/\xi = 0.01$, 0.1, and 0.3 and $\Delta/\xi = 0.5$. The solid lines in Fig. 2 represent the exact formula of $g(\tau)$. Clearly, the numerical computation is in very good agreement with the exact result.

\section{3DEG under RRDA}

We now apply the above algorithm to study 3DEG under RRDA. The uniform 3DEG is a fundamental system in solid-state physics \cite{Kohn} and has been extensively studied for developing the exchange-correlation functional of local density in the frame work of density-functional theory \cite{Hohenberg}. Within the Green's function approach, most of the existing works for studying the system are based on perturbation expansions \cite{Pines,Bohm,Pines2,GM0}. RRDA is considered to be superior to perturbation expansions because it satisfies the microscopic conservation laws \cite{Baym,Baym2}. The similar scheme, the fluctuation-exchange approximation, has been extensively applied to the Hubbard models for studying the mechanism of high-temperature superconductivity in cuprates \cite{Bickers,Pao2,Monthoux,Dahm,Putz,Koikegami,Takimoto,Kontani,Yan}. In different from the Hubbard models that describe narrow-band electrons with short-range Coulomb repulsion, the 3DEG is a system of electrons with infinitive band width and long-range Coulomb interactions. The energy scale of an electron in 3DEG is much larger than that in the Hubbard model and one has to treat the summation over Matsubara frequencies with a much larger cutoff in the numerical computation. Since the Green's function needs to be self-consistently determined by coupled integral equations and the numerical computation is not easy without special method, RRDA has not been applied to 3DEG. Our objects here are to test the efficiency of the present numerical algorithm and to examine the applicability of RRDA to 3DEG. 

\begin{figure} 
\centerline{\epsfig{file=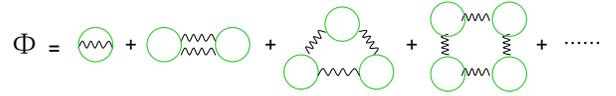,width=8.5cm}}
\caption{(color online) Diagrammatic expressions for `free energy' functional $\Phi$. The line represents the Green's function and the wavy line is the Coulomb interaction.}
\end{figure} 

The three-dimensional electron system with density $n$ at temperature $T$ is embedded in a uniform neutralizing background of positive charge. The Hamiltonian of the system is given by
\begin{eqnarray}
H = \sum_{k\sigma}\epsilon(k)c_{k\sigma}^{\dagger}c_{k\sigma}+\frac{1}{2V}\sum_{kk'q\sigma\sigma'}v(q)c_{k+q\sigma}^{\dagger}c_{k'-q\sigma'}^{\dagger}c_{k'\sigma'}c_{k\sigma} \nonumber
\end{eqnarray}
where $c_{k\sigma}$ annihilates an electron of momentum $k$ and spin $\sigma$, $\epsilon(k) = k^2/2$ is the kinetic energy, $v(q) = 4\pi e^2/q^2$ is the Coulomb interaction, $V$ is the volume of the system, and the term of $q = 0$ is excluded from the summation because of the neutralizing background. Through out the paper, we will use the units in which $\hbar = k_B = m = a = e = 1$ with $m$ the mass of the electron and $a$ the Wigner-Seitz radius. The Coulomb coupling strength is characterized by the parameter
\begin{eqnarray}
r_s = a/a_B    \label{rs}
\end{eqnarray}
with $a_B$ the Bohr radius. The Fermi degeneracy is measured by the ratio $T/E_F$ with $E_F = k_F^2/2$ the Fermi energy and $k_F = (3\pi^2n)^{1/3}$ the Fermi wavenumber.

According to the quantum many-body theory, we start with the electronic Green's function $G(k,i\omega_n)$. It is related to the self-energy $\Sigma(k,i\omega_n)$ via
\begin{eqnarray}
G(k,i\omega_n) =[i\omega_n-\xi_k-\Sigma(k,i\omega_n)]^{-1} \label{G}
\end{eqnarray}
where $\xi_k = \epsilon(k)-\mu$ and $\mu$ is the chemical potential determined by
\begin{eqnarray}
n = \frac{2T}{V}\sum_{kn}G(k,i\omega_n)\exp(i\omega_n\eta) \label {chm}
\end{eqnarray} 
with $\eta$ an infinitesimal small positive constant. With a conserving approximation, the self-energy $\Sigma$ is given as the functional derivative \cite{Baym2}
\begin{eqnarray}
\Sigma =\delta \Phi/\delta G \label {se1}
\end{eqnarray}
where $\Phi$ is the `free energy' functional of the system. The Green's function so determined satisfies the microscopic conservation laws. Under RRDA, $\Phi$ is diagrammatically given by Fig. 3. In space of $(r, \tau)$ with $0 < \tau < \beta$, $\Sigma$ reads
\begin{eqnarray}
\Sigma(r,\tau) = -G(r,\tau)W(r,\tau) \label {se}
\end{eqnarray}
where $W(r,\tau)$ is an effective interaction between electrons as mentioned in the section of Introduction. In space of $(q,i\nu_m)$ with $\nu_m = 2m\pi T$ the bosonic Matsubara frequency, $W$ is expressed as
\begin{eqnarray}
W(q,i\nu_m) = \frac{v(q)}{1-v(q)\chi(q,i\nu_m)} \label {W}
\end{eqnarray}
where $\chi(q,i\nu_m)$ is the bubble as shown in Fig. 3. In terms of $G$, $\chi$ in $(r, \tau)$-space is given by
\begin{eqnarray}
\chi(r,\tau) &=& 2G(r,\tau)G(-r,-\tau) \nonumber\\
&=& -2G(r,\tau)G(r,\beta-\tau) \label{chi}
\end{eqnarray}
where the use of $G(-r,-\tau)=G(r,-\tau)=-G(r,\beta-\tau)$ has been made. The Green's function $G$ is self-consistently determined by Eqs. (\ref{G}) and (\ref{chm}) and (\ref{se})-(\ref{chi}). These integral equations are solved by iterations. Clearly, in each iteration, we need to Fourier transform $G(k,i\omega_n)$ to $G(r,\tau)$ to calculate $\chi(r,\tau)$, transform $\chi(r,\tau)$ to $\chi(q,i\nu_m)$ to obtain $W(q,i\nu_m)$, transform $W(q,i\nu_m)$ to $W(r,\tau)$ to get $\Sigma(r,\tau)$, and finally transform $\Sigma(r,\tau)$ to $\Sigma(k,i\omega_n)$ to return to $G(k,i\omega_n)$.

To numerically do the transforms with guaranteed accuracy, we need to use auxiliary functions. The key points about the transforms and the auxiliary functions are illustrated below.

(i) For transforming $G(k,i\omega_n)$ to $G(r,\tau)$, we choose $G^0(k,i\omega_n) = 1/(i\omega_n-\xi^0_k)$, with $\xi^0_k = \epsilon(k)-\mu_0$ and $\mu_0$ the chemical potential of the non-interacting electron gas, as the auxiliary function as did in the second example given by Eq. (\ref{ex2}). $G^0(k,\tau) = -F(-\xi^0_k)\exp(-\xi^0_k\tau)$ for $0 < \tau < \beta$ has been given in the example. From $G^0(k,\tau)$ to $G^0(r,\tau)$, we need to carefully carry out the integral
\begin{eqnarray}
G^0(r,\tau) &=& -\frac{1}{2\pi^2r}\int^{\infty}_0dkkF(-\xi^0_k)\exp(-\xi^0_k\tau)\sin(kr). \nonumber\\
\label{G0}
\end{eqnarray}
Since it cannot be integrated out analytically, one has to integrate it out numerically. Note that for small $\tau \sim 0$, the factor $kF(-\xi^0_k)\exp(-\xi^0_k\tau)$ decays slowly as $k \to \infty$. In this case, integrating out the slow decaying part analytically, we have
\begin{eqnarray}
G^0(r,\tau) &=& \frac{1}{2\pi^2r}\int^{\infty}_0dkkF(\xi^0_k)\exp(-\xi^0_k\tau)\sin(kr) \nonumber\\
&& -\frac{\exp(\mu_0\tau-r^2/2\tau)}{(2\pi\tau)^{3/2}},\label{G01}
\end{eqnarray}
and the remaining integral is performed numerically with the Filon's rule \cite{Filon}. This formula is useful only for small $\tau$. For large $\tau$, the factor $\exp(\mu_0\tau)$ is large and the rounding error is big. There is the same factor in the integrand in Eq. (\ref{G01}). The formula should be reformed as a summation multiplied by this common factor. However, Eq. (\ref{G0}) is a proper formula for numerical integration for large $\tau$. Having $G^0(r,\tau)$ so obtained, we then need to numerically transform $\delta G(k,i\omega_n) = G(k,i\omega_n)-G^0(k,i\omega_n)$ to $\delta G(r,\tau)$. We select the fermionic Matsubara frequencies using the parameters $[h,L,M] = [2,17,8]$. The obtained results are almost the same as that using the parameters $[h,L,M] = [2,17,12]$. For the momentum $k$ with a cutoff $k_c = 25/a$, its range is separated into four regions: $(0,k_F-\Delta)$ with $\Delta = \min(2T/k_F,k_F/3)$, $(k_F-\Delta,k_F+\Delta)$, $(k_F+\Delta,10/a)$ and $(10/a,k_c)$ and we use 100 uniform meshes in each of them. Since the Fermi distribution is sharp at $k_F$ at low temperature, we need to put dense points there. The $r$ range is divided into three regions $(0,a)$, $(a,10a)$ and $(10a,40a)$ with respectively 50,100,and 150 equal-mesh grids. For the range of $0 < \tau < \beta$, it is divided into 24 segments symmetrically about $\beta/2$. The boundary points of the segments in left of $\beta/2$ are given by 
\begin{eqnarray}
\tau_{12} &=& \beta/2, \nonumber\\
\tau_j &=& \tau_{j+1}/4, ~~~~{\rm for}~j = 1,\cdots,11 \nonumber\\
\tau_0 &=& 0. \nonumber
\end{eqnarray}
In each segment $(\tau_j,\tau_{j+1})$, there are 20 uniform meshes for coordinating the Green's function $G$. It is seen that the meshes become dense as $\tau \to 0$ or $\to \beta$. This choice is necessary at low temperature because the Green's function $G(k,\tau)$ varies dramatically as $\tau \to 0$.

\begin{figure} 
\centerline{\epsfig{file=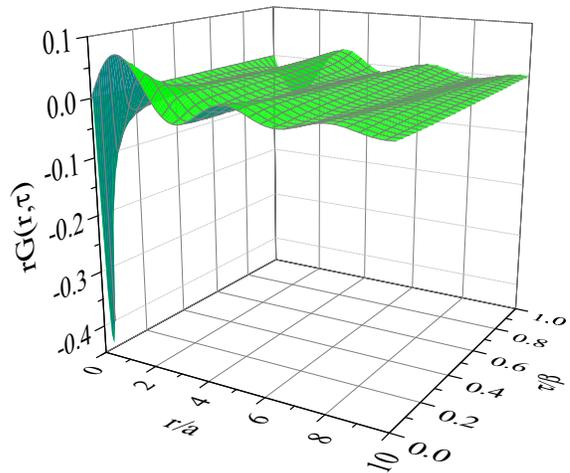,width=8.cm}}
\caption{(color online) Function $rG(r,\tau)$ at $T/E_F =0.05$ and $r_s = 5$.}
\end{figure} 

(ii) With $G^0$, the natural auxiliary function for transforming $\chi(r,\tau)$ to $\chi(q,i\nu_m)$ is chosen as $\chi^0(r,\tau) = -2G^0(r,\tau)G^0(r,\beta-\tau)$, and $\chi^0(q,i\nu_m)$ is given by
\begin{eqnarray}
\chi^0(q,i\nu_m) = -\frac{1}{2\pi^2q}\int^{\infty}_0dk\frac{dF(\xi^0_k)}{dk}J(k,q,\nu_m) 
\label{ch0}
\end{eqnarray}
with 
\begin{eqnarray}
J(k,q,\nu) &=& \frac{1}{2}(k_{+}k_{-}+\nu^2/q^2)\ln\frac{k_-^2+\nu^2/q^2}{k_+^2+\nu^2/q^2}\nonumber\\
&& -kq+\nu(\arctan\frac{qk_-}{\nu}+\arctan\frac{qk_+}{\nu}) \nonumber
\end{eqnarray}
and $k_{\pm} = k\pm q/2$. The remaining transform $\delta\chi(r,\tau) = \chi(r,\tau)-\chi^0(r,\tau)$ to $\delta\chi(q,i\nu_m)$ is carried out numerically using the Filon's rule. 

\begin{figure} 
\centerline{\epsfig{file=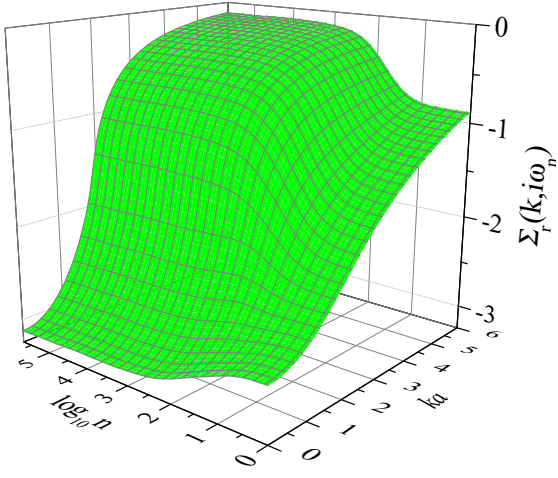,width=8.cm}}
\caption{(color online) Real part self-energy $\Sigma_r(k,i\omega_n)$ at $T/E_F =0.05$ and $r_s = 5$.}
\end{figure} 

(iii) From $W(q,i\nu_m)$ to $W(r,\tau)$, we need numerically transform $\delta W(q,i\nu_m)= W(q,i\nu_m) -v(q)$ to $\delta W(r,\tau)$ while $v(q)$ to $v(r)$ is trivial. At large $q$ or large $\nu_m$, $\delta W(q,i\nu_m)$ becomes $v^2(q)\chi(q,i\nu_m)$. To image the behavior of $\chi(q,i\nu_m)$ in these limits, we consider \begin{eqnarray}
\chi^0(q,i\nu_m) \sim -\frac{3}{2\pi}\frac{\epsilon(q)}{\epsilon^2(q)+\nu^2_m}, ~~{\rm for~} |\epsilon(q)+i\nu_m| \to \infty \nonumber
\end{eqnarray}
as a measure of it. The transform of $\epsilon(q)/[\epsilon^2(q)+\nu^2_m]$ from $\nu_m$ space to $\tau$ space is the same as in the first example given by Eq. (\ref{ex1}) by noting $p = \epsilon(q)/T$. For large $q$ and low $T$, $\epsilon(q)/T$ can be very large. Therefore, the cutoff for $\nu_m$ should be large enough. We use the parameters $[h,L,M] = [2,22,8]$ for selecting $\nu_m$'s (giving rise to almost the same results as that of $M = 12$). The cutoff $\nu_m$ is $\nu_c = 2^{25}\pi T \approx 3.3\times 10^7\pi T$. In our calculation, we first transform $\delta W(q,i\nu_m)$ to $\delta W(r,i\nu_m)$ choosing $\delta W^0(q,i\nu_m) = v^2(q)\chi(0,i\nu_m)/[1-v(q)\chi(0,i\nu_m)]$ as the auxiliary function. $\delta W^0(r,i\nu_m)$ is given by
\begin{eqnarray}
\delta W^0(r,i\nu_m) =[\exp(-q_mr)-1]/r \nonumber
\end{eqnarray}
with $q_m = \sqrt{-4\pi\chi(0,i\nu_m)}$. The $q$ integral in the numerical transform is performed with the Filon's rule using 100, 200, and 100 uniform meshes in ranges $(0, 1/a)$, $(1/a, 10/a)$ and $(10/a, 35/a)$, respectively. Finally, we transform $\delta W(r,i\nu_m)$ to $\delta W(r,\tau)$. 

\begin{figure} 
\centerline{\epsfig{file=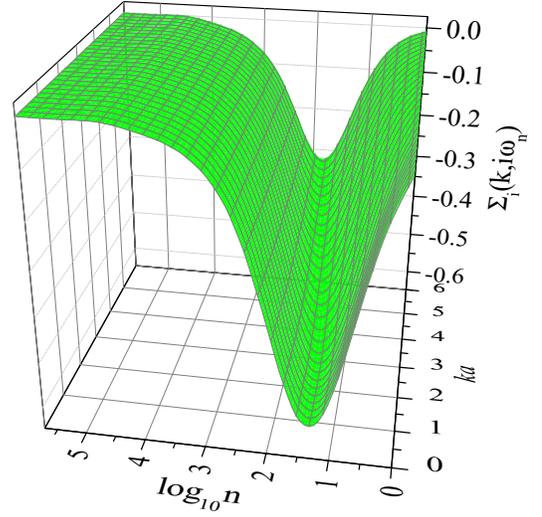,width=7.5cm}}
\caption{(color online) Imaginary part self-energy $\Sigma_r(k,i\omega_n)$ at $T/E_F =0.05$ and $r_s = 5$.}
\end{figure} 

(iv) The self-energy is separated to the Fock (that is independent of $\omega_n$) and the remaining terms. The two terms are transformed separately from $(r,\tau)$ space to $(k,i\omega_n)$ space. 

Note here that under RRDA $\chi(0,i\nu_m)$ is not zero and is different from RPA. To see it, by inserting the RPA self-energy into the Green's function and calculating $\chi(q,i\nu_m)$, one then finds $\chi(0,i\nu_m) \ne 0$. RRDA is such a process that the Green's function is corrected again and again until the self-consistency is satisfied. A related problem is the plasmon excitation in the system. With RPA, the frequency of plasmon is determined by the singularity of the summation of ring diagrams. Under RRDA, however, it should be determined by the singularity of a two particle propagator. The effective particle-hole interaction in the two-particle propagator is determined by second functional derivative of $\Phi$ with respect to the single-particle Green's function $G$. Therefore, the diagram of the irreducible two-particle propagator is not a simple bubble.

The maximum value of $\epsilon(q)$ here takes a role of criterion in determining the cutoff $\nu_c$ of Matsubara frequency $\nu_m$. The largest $q$ is $35/a$, leading to largest $\epsilon(q) = 0.27\times 35^2E_F \approx 333E_F$. For $T/E_F = 0.01$, we have $\nu_c \approx 10^6 E_F >> \epsilon(q)$. In a narrow-band system such as Hubbard model, instead of $\epsilon(q)$ as appeared above, we may use the band width to estimate the cutoff. To see this, we start from the more general expression for $\chi_0(q,i\nu_m)$ 
\begin{eqnarray}
\chi^0(q,i\nu_m) = \frac{4}{V}\sum_k\frac{\Delta(k,q)[F(\xi^0_k)-F(\xi^0_{k+q})]}{\nu_m^2+\Delta^2(k,q)} 
\nonumber
\end{eqnarray}
with $\Delta(k,q) = \epsilon(k)-\epsilon(k+q)$. Clearly, $\nu_c$ should be much larger than the maximum value of $\Delta(k,q)$. In the Hubbard model, the magnitude of the later is in the order of the band width. 

The points in the $(k,i\omega_n)$ or $(q,i\nu_m)$ and $(r,\tau)$ spaces may not be necessarily chosen so dense to coordinate the functions. Even with so many points, the computation time for solving the integral equations of the Green's function with a personal microcomputer is only a few seconds, which is tolerable to us.

\begin{figure} 
\centerline{\epsfig{file=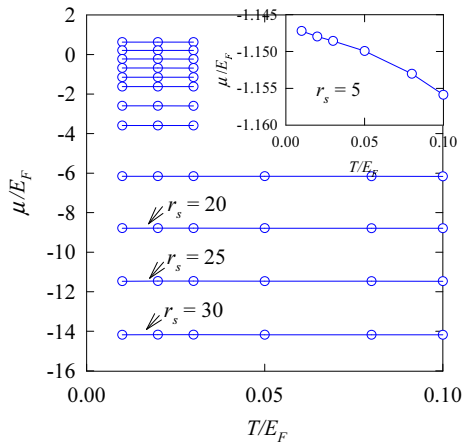,width=6.5cm}}
\caption{(color online) Chemical potential $\mu$ as function of temperature $T$ for parameters, from top, $r_s = $1, 2, 3, 4, 5, 6, 8, 10, 15, 20, 25, 30. The inset is a zoom in $\mu-T$ picture at $r_s = 5$.}
\end{figure} 

With the techniques given above, we have solved the integral equations. Fig. 4 shows the function $rG(r,\tau)$ at $T/E_F = 0.05$ and $r_s = 5$. This function varies dramatically in a region close to origin (0,0) where there is a sharp dip in the surface given by $rG(r,\tau)$. This behavior can be seen from part of the free-particle Green's function $G^0(r,\tau)$ as given by the last term in Eq. (\ref{G01}). At large $r$, the surface seems like a wave. This wave is related to the Friedel oscillations as seen from Eq. (\ref{G01}) where the Fermi distribution function varies drastically at $k_F$. In Figs. 5 and 6, for the same parameters $T/E_F = 0.05$ and $r_s = 5$, we show the real part and imaginary part of the self-energy $\Sigma(k,i\omega_n)$, respectively. As $\omega_n \to \infty$, $\Sigma(k,i\omega_n)$ goes to the Fock exchange that is real as shown in Fig. 5. At $ka \sim 2$, $\Sigma_r(k,i\omega_n)|_{n\to\infty}$ varies dramatically, showing a logarithmic behavior due to the Coulomb interaction. On the other hand, $\Sigma_i(k,i\omega_n)|_{n\to\infty}$ becomes zero as shown by Fig. 6. In the limit $k \to \infty$, $\Sigma(k,i\omega_n)$ vanishes.

Figure 7 shows the chemical potential $\mu$ as a function of temperature $T$ at various coupling constant $r_s$. At each $r_s$, $\mu$ seems as a constant at low temperature. Actually, $\mu$ slightly decreases with $T$ as shown in the inset of Fig. 7 for $r_s = 5$, which is a general feature of $\mu$ because electrons occupy higher energy levels with larger density of states due to the thermal excitations. 

\begin{figure} 
\centerline{\epsfig{file=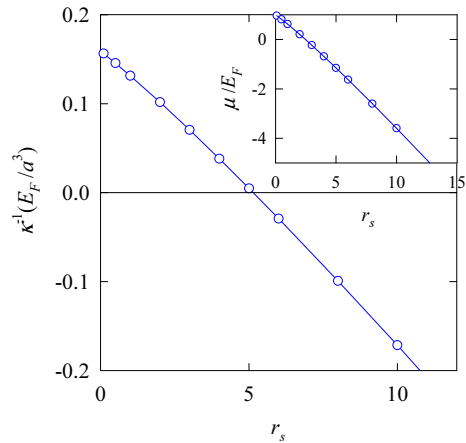,width=6.5cm}}
\caption{(color online) Inverse compressibility $\kappa^{-1}$ as function of $r_s$ at $T/E_F = 0.01$. The inset shows $\mu$ as function $r_s$ at $T/E_F = 0.01$.}
\end{figure} 

Shown in Fig. 8 is the inverse compressibility defined by
\begin{eqnarray}
\kappa^{-1} = n^2(\frac{\partial\mu}{\partial n})_T = -\frac{nr_s}{3}(\frac{\partial\mu}{\partial r_s})_T. \label{cmp}
\end{eqnarray}
The corresponding $\mu$ as a function $r_s$ is depicted in the inset. At small $r_s$ (high density), $\kappa^{-1}$ is positive, implying that the system is stable. While at large $r_s$, $\kappa^{-1}$ becomes negative and the system is unstable. The critical value is $r_s \approx 5$ where $\kappa$ goes to infinitive, implying that Wigner crystallization may takes place in the system. This critical value $r_s \approx 5$ under RRDA drops in the range $4.83 \le r_s \le 104$ estimated by the earlier works \cite{Nozieres,Coldwell,Wette,Horn,Glyde,Ceperley2,Utsumi} with 4.83 as the prediction of the Hartree-Fock perturbation \cite{Fetter}. The earlier MC result \cite{Ceperley} for the critical value of Wigner crystallization is $r_s = 67$.  

We here need to emphasize that the compressibility $\kappa$ is not equal to $-\chi(0,0)/n^2$. According to the compressibility sum rule, $\kappa$ is related to an irreducible two-particle propagator. As mentioned above, the diagram of the irreducible two-particle propagator is not a simple bubble under RRDA. 

\begin{figure} 
\centerline{\epsfig{file=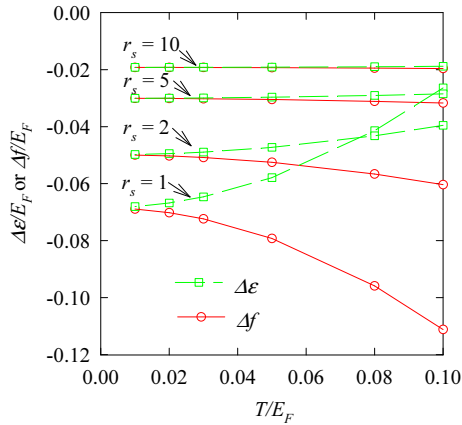,width=6.5cm}}
\caption{(color online) Energy $\Delta \epsilon$ and free energy $\Delta f$ per particle as functions of $T$ at $r_s = 1$, 2, 5, and 10.}
\end{figure} 

With the results for the Green's function and the self energy, we calculate the energy $E = \langle H\rangle$. $E$ is given by 
\begin{eqnarray}
E &=& \sum_{k}[2\epsilon(k)+\Sigma_x(k)]n(k)\nonumber\\
 && -\frac{T}{2}\sum_{qm}\frac{v^2(q)\chi^2(q,i\nu_m)}{1-v(q)\chi(q,i\nu_m)}   \label{E}
\end{eqnarray}
where $\Sigma_x(k)$ is the Fock exchange part of the self-energy $\Sigma_x(k) = \Sigma(k,i\infty)$, and $n(k)$ is the distribution function. On the other hand, $E$ can be also obtained from the thermodynamic function $\Omega$ as 
\begin{eqnarray}
E = \frac{\partial}{\partial\beta}(\beta\Omega)_{\mu V}+\mu N \label{E2}
\end{eqnarray}
where $N = nV$ is the total number of the electrons and $\Omega$ is given by \cite{Luttinger}
\begin{eqnarray}
\Omega = \{\Phi -\sum_{k,\sigma,n}\exp(i\omega_n\eta)[\Sigma G - \ln(-G)]\}/\beta.   \label{tp}
\end{eqnarray}
Under the conserving approximation, because of Eq. (\ref{se1}), the two equations (\ref{E}) and (\ref{E2}) are equivalent. Having $\Omega$, we obtain the free energy $F$,
\begin{eqnarray}
F = \Omega+\mu N, \label{fe}
\end{eqnarray}
which is related to $E$ by $F = E - TS$ with $S$ as the entropy. In Fig. 9, we plot the results for $\Delta\epsilon \equiv (E-E_0)/N$ and $\Delta f \equiv (F-E_0)/N$ as functions of $T$ with 
\begin{eqnarray}
E_0/N = \frac{1.105}{r_s^2}-\frac{0.458}{r_s}~~{\rm (a.u.)}  \label{E0}
\end{eqnarray}
as the ground-state energy in atomic unit (a.u.) given by the Hartree-Fock perturbation \cite{Fetter}. $E$ increases with $T$ because of the thermal excitations. But the increment is less than the heat $TS$, so $F$ decreases with $T$. In the limit $T \to 0$, both of them become the same result, the ground state energy. $\Delta\epsilon$ and $\Delta f$ at $T = 0$ are the ground-state correlation energy per particle $\epsilon_c$. Shown in Fig. 10 is the result for $\epsilon_c$ as a function of $r_s$. The present calculation (RRDA) is compared with MC \cite{Ceperley} and RPA. By RPA, $\epsilon_c$ is given by \cite{Fetter}
\begin{eqnarray}
\epsilon_c^{RPA} &=& \frac{1}{2N}\sum_{q}\int_{-\infty}^{\infty}\frac{d\nu}{2\pi}\{\ln[1-v(q)\chi^0(q,i\nu)]\nonumber\\
&&~+v(q)\chi^0(q,i\nu)\}.  \label{ERPA}
\end{eqnarray}
Clearly, in the whole range of the coupling constant plotted here, the present RRDA calculation is much closer to MC than RPA. The present result is even almost the same as the MC simulation at $r_s \ge 2$.  Note that the result from Eq. (\ref{E}) where $\Sigma_x$, $n(k)$ and $\chi$ are replaced with the corresponding functions of free electrons as in RPA is not the same as from Eq. (\ref{ERPA}) plus $E_0/N$ because RPA is not a conserving approximation. 

\begin{figure} 
\centerline{\epsfig{file=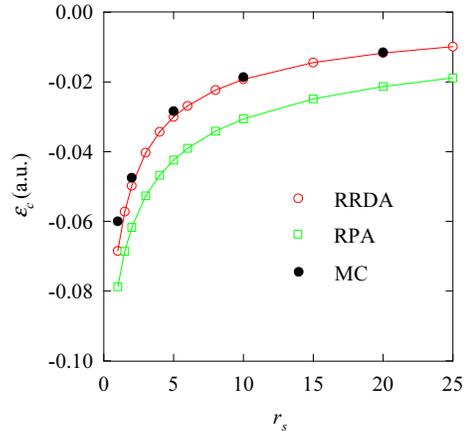,width=6.5cm}}
\caption{(color online) Ground-state correlation energy $\epsilon_c$ (in atomic unit) per particle as functions of $r_s$. The present calculation (circles) is compared with RPA (squares) and MC (solid circles). }
\end{figure} 

In our previous work \cite{Yan2}, we have studied the two-dimensional interacting electron gas (2DEG) under RRDA. In that time, we could not perform approximated discrete Fourier transform from the Matsubara frequency axis to the imaginary time axis, but carried out the calculation using approximated series summation algorithm for the direct summation over the Matsubara frequency. To test the present numerical method of the discrete Fourier transform, we have reinvestigated the 2DEG system. For solving the integral equations determining the Green's function, the numerical computation with the present algorithm is much easier and faster than that with the previous method. Shown in Fig. 11 is the ground-state energy as a function of the coupling constant $r_s$ for 2DEG. The present calculation (circles) reproduces precisely the previous results (diamonds). For comparison, the results of MC simulation (solid circles) \cite{Attaccalite} and RPA (squares) are also depicted in Fig. 11. We here make a correction for the RPA result in the previous work. The previous RPA notation for the ground-state energy is not obtained from Eq. (\ref{ERPA}) but from (\ref{E}). That is not the usual meaning for the ground-state energy of the RPA calculation. 

As seen from Figs. 10 and 11, RRDA reproduces quite accurately the ground-state energy of the MC results, better for higher dimensional system. A question then raises: Why the critical value $r_s \approx 5$ for the singularity of compressibility of 3DEG is very different from the MC value 26 for the ferromagnetization or 67 for the Wigner crystallization? To answer it, we recall the expression $\mu = (\partial E/\partial N)_{V,T=0}$ and write $\kappa^{-1}$ for $T = 0$ as 
\begin{eqnarray}
\kappa^{-1} = \frac{n^2}{V}(\frac{\partial^2E}{\partial n^2})_{V,T=0}. \label{kappa}
\end{eqnarray}
Clearly, though the ground-state energy is obtained accurately with an approximation, its second derivative with respect to $n$ may in general not be good. Under RRDA, the summation of the most divergent ring diagrams gives the predominate contribution to the ground-state energy. However, to its derivatives, the contribution from the neglected diagrams may be also significant. 

\begin{figure} 
\centerline{\epsfig{file=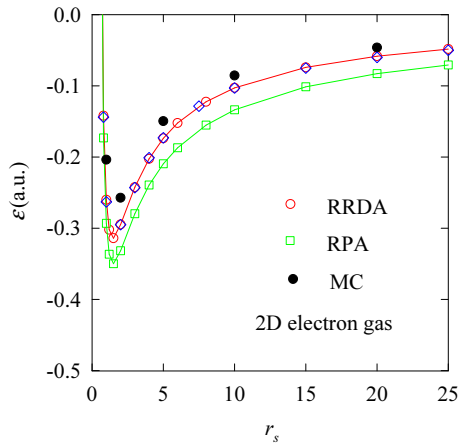,width=6.5cm}}
\caption{(color online) Ground-state energy $\epsilon$ (in atomic unit) per particle as functions of $r_s$ for two-dimensional electron gas. The present calculation (circles) is compared with RPA (squares) and MC (solid circles). The diamonds are the RRDA results of the previous numerical calculation.}
\end{figure} 

Having obtained the total energy $E$ and the free energy $F$, we then get the entropy $S$ as
\begin{eqnarray}
S = (E-F)/T.  \label{entpy}
\end{eqnarray}
Shown in Fig. 12 is $s = S/N$ as a function of $T$ at various coupling constant $r_s$. The temperature in Fig. 12 is normalized with the energy $e^2/a$ (that is a scale larger than $E_F$ for $r_s > 1.84$). As is seen, for smaller $r_s$, $s$ varies linearly in a wider range of low temperature. At large $r_s$, the tangent of $s$ decreases as $T$ increasing, reflecting the strong coupling effect. The low temperature result for $s$ at large $r_s$ is hard to calculate accurately since the difference between $E$ and $F$ is very small. The situation has been shown in Fig. 9 where $T$ is normalized with $E_F$ that is a smaller scale than $e^2/a$ for large $r_s$. The overall relative error of the numerical calculation for $E$ and $F$ is in the range $(0.001,0.01)$. The difference between them can fall into the error bars at large $r_s$ and at low $T$. The numerical calculation for $s$ is therefore meaningful only at high temperature. At large $r_s$, the electrons are strongly coupled. The average energy of an electron is in the order of $e^2/a$. Therefore, $e^2/a$ is the proper energy scale for strong coupling case. 

With the results for entropy $S$, we calculate the specific heat $C$ defined as
\begin{eqnarray}
C = T(\frac{\partial S}{\partial T})_n.  \label{spcfh}
\end{eqnarray}
In Fig. 13, we plot the result for $C$ as a function of $r_s$ at various $T$. $C$ is normalized by $C^F$ the specific heat of the free electrons, 
\begin{eqnarray}
C^F = \pi^2NT/2E_F.  \label{fspcfh}
\end{eqnarray}
As seen from Fig. 13, $C/C^F$ monotonically decreases with $r_s$ for $r_s > 1$. For fixed $r_s$, $C/C^F$ is smaller at higher $T$ reflecting the tangent change of $s(T)$ as shown in Fig. 12. The dashed line in Fig. 13 represents the Gell-Mann high-density expansion for the specific heat \cite{GM}
\begin{eqnarray}
C^{GM}/C^F = \{1+\alpha r_s[\log(\pi/\alpha r_s)-2]/2\pi\}^{-1}  \label{hdx}
\end{eqnarray}
with $\alpha = (4/9\pi)^{1/3}$. $C^{GM}/C^F$ first decreases from unity as $r_s$ departing from 0 and then turns to increases with $r_s$. Of course, it is valid at small $r_s$. As $r_s \to 0$, the result given by RRDA should be close to this expansion. 

\begin{figure} 
\centerline{\epsfig{file=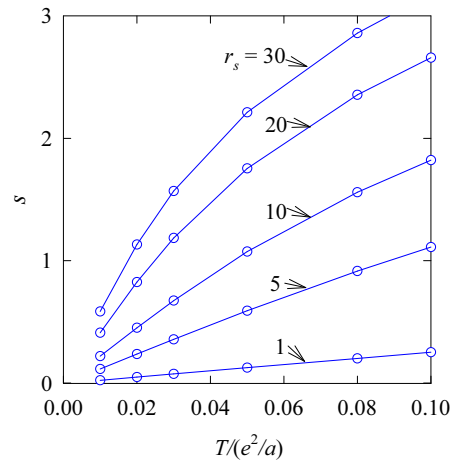,width=6.3cm}}
\caption{(color online) Entropy $s$ per particle as functions of $T$ at $r_s$ = 1, 5, 10, 20, and 30. }
\end{figure} 

\section{Conclusion}

By conclusion, we have developed the approximated algorithm for the discrete Fourier transform. When the function to be transformed is piecewise smooth, its transformation can be accurately obtained using a number of selected points with corresponding weights. This algorithm reduces the requirement for computer memory storage and enhances the numerical computation efficiency by several orders. Its accuracy has been examined by examples.

We have applied the numerical algorithm to study the three-dimensional interacting electron gas under the renormalized-ring diagram approximation. The integral equations determining the Green's function are easily solved by the present numerical algorithm. Since the band width of the system is infinitive, the number corresponding to the cutoff of the fermionic or bosonic Matsubara frequency at low temperature in the numerical computation is extremely large. With the present algorithm, instead of computing the functions at about $2^LM$ Matsubara frequencies in each iteration, one needs to calculate them at only $LM+1$ selected ones. The parameters $L$ and $M$ used here are $L = 17$ (22) for fermions (bosons) and $M =8$. The requirement for the computer memory storage is greatly reduced and the efficiency is $c = 2^L/L \approx 7.7\times 10^3$ ($1.9\times 10^5$) for $L = 17$ (22). We have obtained the results for the chemical potential, compressibility, free energy, entropy, and specific heat of the system. The ground-state energy obtained by the present calculation is in very good agreement with the result of Monte Carlo simulation.  

\begin{figure} 
\centerline{\epsfig{file=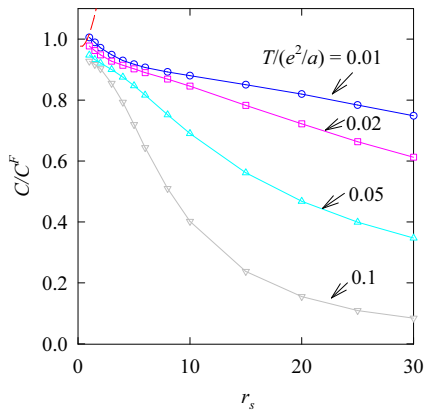,width=6.cm}}
\caption{(color online) Specific heat $C$ normalized by $C^F$ that of free electrons as functions of $r_s$ at $T/(e^2/a)$ = 0.01, 0.02, 0.05, and 0.1. The dashed line at small $r_s$ is the Gell-Mann high-density expansion \cite{GM}.}
\end{figure} 

\acknowledgments

This work was supported by the National Basic Research 973 Program of China under Grant No. 2011CB932700, NSFC under Grants No. 10774171 and No. 10834011, and financial support from the Chinese Academy of Sciences for advanced research.

\end{document}